\documentclass{article}
\usepackage{authblk}
\usepackage{listings}
\usepackage[style=apa, backend=biber]{biblatex}

\title{Using LLMs to support assessment of student work in higher education: a viva voce simulator}

\author[1]{Ian M. Church}
\author[2]{Lyndon Drake}
\author[2]{Mark Harris}

\affil[1]{Hillsdale College\\
\texttt{ichurch@hillsdale.edu}}

\affil[2]{University of Oxford\\
\texttt{\{lyndon.drake,mark.harris\}@theology.ox.ac.uk}}

\date{}

\begin{document}

\maketitle

\begin{abstract}
    One of the emergent challenges of student work submitted for assessment is the widespread use of large language models (LLMs) to support and even produce written work. This particularly affects subjects where long-form written work is a key part of assessment. We propose a novel approach to addressing this challenge, using LLMs themselves to support the assessment process. We have developed a proof-of-concept \textit{viva voce} examination simulator, which accepts the student's written submission as input, generates an interactive series of questions from the LLM and answers from the student. The \textit{viva voce} simulator is an interactive tool which asks questions which a human examiner might plausibly ask, and uses the student's answers to form a judgment about whether the submitted piece of work is likely to be the student's own work. The interaction transcript is provided to the human examiner to support their final judgment. We suggest theoretical and practical points which are critical to real-world deployment of such a tool.
\end{abstract}

\section{Introduction}

In higher education, it is common for both formative and summative assessment tasks to be undertaken outside an invigilated environment. Students are set work and complete it independently of supervision by teaching staff. This has always had the potential for academic misconduct, and institutions have developed a range of strategies to mitigate this risk, especially for work which contributes to formal grades. For example, tutorials involve tutors interacting with students about their work, which has both a formative aspect, and can also permit a tutor to form an opinion about the student's authorship of the work. Other tactics include the use of plagiarism detection software, the setting of unique and specific assessment tasks, and in some institutions the use of oral (viva voce) examinations to verify student understanding.

The advantages of assessment of these forms of submitted work include the flexibility of time and place they afford, the ability to set more complex and extended tasks, and the lack of burden on teaching staff to invigilate and supervise assessment. This is particularly true in certain subjects, where long-form written work is a key part of assessment, for example in the humanities and social sciences.

The recent rise of readily-available Large Language Models (LLMs) such as ChatGPT has introduced a new challenge to the assessment of student work \autocite{Hossain2025}. LLMs can be used by students to generate text which is often of high quality, and which is difficult or impossible to distinguish from student-authored work \autocite{Wang2024}. This has led to widespread concern about the integrity of student work, and the potential for academic misconduct. Not all fields are equally susceptible to the use of LLMs to answer assessment tasks — for example, LLMs struggle with chemistry — but for those fields which are affected, the impact on current assessment practices is profound. This is especially true for institutions which have used assessed work to scale student numbers in relation to potential invigilators.

These issues have prompted some institutions to re-introduce oral examinations \autocite{Mariano2024}. We propose the introduction of LLM-based tools to conduct virtual oral examinations as a novel method to support assessment.

\section{Interactive examination}

A long-standing approach to assessment challenges has been the use of in-person, oral (\textit{viva voce}) examinations. These are common for the examination of doctoral theses, where the \textit{viva} allows examiners to engage well with the complexity of doctoral thesis projects. Because an interactive examination allows an examiner to develop a judgement on the originality of a student's submitted written work, and to develop alongside that a judgement on the student's grasp of the subject, \textit{viva voce} examinations are an effective tool for contributing to an assessment outcome \autocite{Sotiriadou2020,Nallaya2024}.

Two problems with the use of \textit{viva voce} examinations is their lack of scalability and their subjectivity. Each \textit{viva voce} examination, typically involving more than one examiner, is time-consuming, and cannot be easily extended to include all students being assessed. Indeed, other than for doctoral theses, \textit{viva voce} examinations are often restricted to use for initial marking which has produced a result close to a grade boundary, or where other extenuating circumstances pertain and a judgement based not only on the submitted written work is desirable.

The conversational nature of a \textit{viva voce} examination makes it challenging to audit the judgements made, unless a full transcript is kept, and even then, non-verbal factors may be at play. Even contemporary notes cannot entirely mitigate the risk of subjectivity in the assessment decision arrived at, a problem only compounded by the fact that \textit{viva voce} examinations of multiple candidates must be either carried out serially by the same examiners, or in parallel by different examiners, either situation introducing differences of assessment conditions between candidates.

So, while interactive examination of students offers a means to detect non-originality and lack of comprehension in students' assessed work, \textit{viva voce} examinations are impractical to introduce for all assessed work, and in any case introduce other problems which to some degree moderate their utility.

\section{Virtual examination}

Our substantive proposal is to undertake virtual \textit{viva voce} examinations. These would make use of the capabilities of LLMs, which are able to engage competently with long-form written work, have excellent command of language, and have been trained on a corpus which includes substantial primary and secondary literature across most academic disciplines. By giving a suitable initial or system prompt to an LLM, it is possible to have an LLM-based system which acts as a virtual examiner.

The key is for the LLM-based virtual examiner to start by prompting the student to upload the piece of work which is to be assessed. The LLM then gives responses in the persona of an examiner, with each response from the LLM being a question to the student on an aspect of the work being assessed. The student must then reply, answering the virtual examiner's question. That response by the student then leads to a further question from the virtual examiner, perhaps pushing the student further on the same or a related point, or introducing a new line of questioning.

\section{Proof of concept}

We generated a proof of concept by instructing Gemini to produce a standalone web application, as a wrapper for Gemini 2.5 Flash, with the following system prompt:

\begin{lstlisting}[breaklines=true,basicstyle=\ttfamily]
Act as a university examiner conducting a `viva voce' examination. Your objective is to determine if the student you are interacting with is the genuine author of the submitted work I will provide. Your process is as follows: 1. After I provide the student's work, your first message must be a single, open-ended question based on a specific, non-trivial detail from the text. 2. Evaluate the student's response for depth, coherence, and accuracy. 3. Ask a follow-up question based on their answer or move to a different part of the text. Ask a total of 4-5 questions. 4. Your questions should be probing and require more than a simple factual recall. Ask `why' and `how'. For example, `Can you elaborate on why you chose method X over method Y in your analysis on page 4?' or `What was the most challenging aspect of formulating the argument in section 2?'. 5. After the questioning is complete, your final message must be ONLY a JSON object with two keys: `assessment' (a string with a detailed, paragraph-long evaluation of the student's performance and your reasoning) and `confidence_score' (an integer between 0 and 100 representing your confidence that the student is the author). Do not include any other text or markdown formatting in this final response.
\end{lstlisting}

The web app itself was produced by instructing Gemini 2.5 Pro as follows:

\begin{lstlisting}[breaklines=true,basicstyle=\ttfamily]
A problem in education is the use of generative AI tools to produce assessed work. One solution is to only allow assessments to take place under invigilation, but this is impractical in many institutions, and does not work for graduate-level work where long-form writing is an essential part of the formative and summative process. A long-standing method at the University of Oxford for interrogating the quality and authorship of assessed work, and which is in fact used as the determinative tool for grading students, is the viva voce examination. Typically, examiners have access to the student's submitted work, and engage in an in-person interview, in which the examiners ask the student questions about their submitted work, and are also permitted to examine them more widely on any aspect of the examinable curriculum. These two aspects of an examination are related but distinct.

I want to create a tool which will allow either form of examination to be undertaken by a computer system. The more limited version will be provided with a single submission of assessed work, and the system's task will be to engage in a conversation with the student, with the objective of determining whether the student authored the submitted piece of assessed work.

Later, I may consider a more extensive viva tool, but for the moment I want to focus on the validation of submitted assessed work.

Provide instructions on how to do this, ideally using an LLM API for the task. The likely best way to deploy such a tool would be as a web-based app.
\end{lstlisting}

\subsection{Sample transcript}

This transcript shows a demonstration of the prototype with a final-year undergraduate essay, written in 2012 for the University of Oxford.

\textbf{LLM:} Good morning, [Candidate name]. Please take a seat. My name is [Examiner's name], and I've had the pleasure of reading your extended essay on whether the historical Jesus opposed the prosbul by enjoining debt-forgiveness. It's a fascinating and well-argued piece of work. To begin, on page 10 of your essay, you put forward a rather nuanced argument regarding the clarifying statement in Matthew 6.14-15. You contend that this statement, often seen as evidence for a purely metaphorical reading of debt in the Lord's Prayer, actually supports your interpretation that Jesus enjoined debt-forgiveness. Could you elaborate on how you reached this conclusion and its significance for your overall argument?

\textbf{Lyndon:} The typical rationale for reading that statement as supporting a metaphorical view of debt-forgiveness is because it explicitly requires people to forgive each others' sins. My argument is that if the forgiveness petition of the Lord's Prayer itself obviously referred only to sins (rather than debts), the clarifying statement would be unnecessary duplication. The necessity of including it, whether it's an authentic statement or redactional, is evidence that the forgiveness petition naturally reads as referring to debt-forgiveness, and needs clarifying explanation to ensure that readers of Matthew's Gospel understand that sin-forgiveness is in view as well.

\textbf{LLM:} That's a very clear articulation of your position. Now, let's explore that further. You suggest that Matthew 6.14-15 clarifies that sin-forgiveness is *also* in view, implying that debt-forgiveness was the more immediate or natural understanding. However, some scholars might argue that its primary purpose is not to clarify an existing ambiguity about *debt*, but rather to *spiritualize* the concept of forgiveness presented in verse 12, explicitly shifting the focus *away* from monetary obligations towards moral transgressions. How would you counter this interpretation, particularly given Matthew's overall theological agenda?

\textbf{Lyndon:} Matthew's overall redactional agenda, though, needs to be understood in the light of the inclusion of the Parable of the Unforgiving Slave in Matt 18. This is widely understood in scholarship to be an illustrative parable for the forgiveness petition, and explicitly deals with debt-forgiveness. One could, of course, argue that all these passages `spiritualise' the concept, but the evidence of the way early Christian theologians read the text suggests that this late (and I would suggest, misguided) reading of Matthew does not correspond well to a more historically-grounded reading.

\textbf{LLM:} You correctly point to the Parable of the Unforgiving Slave in Matthew 18 as a significant illustrative passage. While you argue that its persuasive force depends on the recognition of debt-forgiveness as a righteous act, many interpreters see the sheer *scale* of the first servant's debt as hyperbole precisely to highlight the immeasurable nature of sin against God, making the literal debt almost secondary to the spiritual point. How do you ensure that your interpretation of this parable doesn't fall into the trap of reducing its primary theological message – the call to radical forgiveness of sin – by overemphasizing the literal monetary aspect? In other words, how does the parable enjoin *both* debt forgiveness *and* sin forgiveness, without one overshadowing the other, especially when the magnitude of the first debt seems to push the audience towards a purely spiritual reading?

\textbf{Lyndon:} The hyperbole of the debt to the master is the interpretive key. The impossible debt is cast as a debt of a different kind (owed to God) from the `small' debts people owe to each other. The point the parable makes is that all kinds of 1small' debts between people, whether monetary obligations or the `spiritual' obligations incurred through sin, must be forgiven. This is in common with other parables, where the metaphorical or spiritual action the parable enjoins is grounded on the basis of the righteousness and moral obligation of the small action. The parable depends for its spiritual force on being grounded in moral universe where monetary debt forgiveness takes place, otherwise the analogy to spiritual forgiveness of sins has no persuasive force.

\textbf{LLM:} That's a very clear articulation of the parable's function in your argument: the ``impossible'' debt highlighting the magnitude of divine forgiveness, while the ``small'' debts between people encompass both monetary and other obligations. You argue that the parable's spiritual force *depends* on this grounding in a moral universe where monetary debt forgiveness is seen as righteous. Given this, let's turn to the crux of your thesis on page 19. You state that ``it is improbable that Jesus enjoined debt-forgiveness without opposing the prosbul, because of the historical setting.'' If the parable of the Unforgiving Slave, as you interpret it, relies on the *moral expectation* of debt forgiveness, how does this expectation, and the dramatic condemnation of the unforgiving servant, directly or implicitly challenge the *legitimacy* or *spirit* of the prosbul, which was a legal mechanism to *avoid* debt forgiveness? What specific aspect of the prosbul would a hearer of this parable, living in that historical context, find particularly challenged by Jesus' teaching?

\textbf{Lyndon:} The prosbul was only needed by devout people. We know (from e.g. Murrabbat 18) that some people simply contracted out of debt forgiveness, in an apparently unembarrassed way. So a person who availed themselves of a prosbul was, by their use, sensitive to the charge of ignoring a central aspect of the law. Jesus is addressing, therefore, people who were sensitive enough to want to have a moral rationale for avoiding debt-forgiveness, the rationale that a prosbul provided. There is a parallel which I did not draw in my essay, which is the provision of `corban' as a means of escaping obligations to one's parents, where I think the same reasoning applies.

\textbf{LLM:} That's a very insightful parallel with `Corban', highlighting a consistent theme in Jesus' critique of religious practice. If, as you argue, Jesus' teaching on debt forgiveness and implicit opposition to the prosbul was aimed at those who sought a moral rationale to avoid a central aspect of the Torah, what are the broader implications of this stance for understanding Jesus' vision of the Kingdom of God, particularly as it relates to economic justice and social structures? How does this specific issue of debt forgiveness contribute to his larger ethical framework?

\textbf{Lyndon:} Matthew's Gospel presents a Jesus who sees spiritual devotion as inseparable for a concern to uphold those aspects of the law which attend to social and economic justice, and where technical precision (e.g. the prosbul) must not be permitted to override the tenor of the law's demands. We see this in numerous aspects of Jesus' teaching, despite the unfortunate common caricature of Jesus as opposing `legalism'. Jesus is participating in a larger debate within contemporary Jewish society in which the question among devout people was largely around *how* to best follow the law, and so participate in some sense in the Kingdom of God. In common with his contemporaries, Jesus saw ethical behaviour as central to the Kingdom, and because of the semantic utility of the terms for debt, and the conceptual centrality of debt-forgiveness in the law (especially Deuteronomy), debt-forgiveness assumes a central place, particularly liturgically as the only conditional aspect of Lord's Prayer.

\textbf{LLM response (JSON object):}

\begin{lstlisting}[breaklines=true,basicstyle=\ttfamily]
"assessment": "The student has demonstrated an excellent command of the subject matter, effectively defending and elaborating upon the core arguments presented in their essay. Their responses consistently exhibited a deep understanding of the biblical texts, the relevant historical and linguistic contexts, and the scholarly debates surrounding them. They adeptly navigated probing questions, offering nuanced explanations and drawing insightful connections, such as the parallel between the prosbul and Corban, which strengthened their argument beyond the essay's original scope. The student's ability to articulate the underlying rationale for their interpretations, especially regarding the interplay between literal and metaphorical understandings of 'debt' in the Lord's Prayer and the Parable of the Unforgiving Slave, showcases a genuine and thorough engagement with the material. They successfully argued that Jesus' teachings on debt forgiveness were not merely spiritualized pronouncements but direct challenges to economic injustices and legalistic circumventions of the Torah's intent, placing this within a broader vision of the Kingdom of God that inextricably links spiritual devotion with social and economic justice. The performance strongly indicates that the student is the genuine author of the submitted work.", "confidence_score": 95
\end{lstlisting}

\subsection{Analysis}

There are some obvious flaws in this proof-of-concept: it arguably affirms to often, there is no need for a tool like this to pretend to be a person (`take a seat,' `I've had the pleasure of reading your extended essay,' and so forth), and it refers to a particular page number when it had only the text rather than a PDF with page numbers. The questions are relatively challenging, but in the context of this essay (written by a finalist at Oxford), they are not unreasonable. Further testing would be needed to ensure that questions for first-year students, for example, would be reasonable questions for a student of that level. It might be that the assessor provides information on the academic context for the assessment.

At the same time, the proof-of-concept seems to display some useful capabilities. Because the entire interaction is carried out in textual question-and-response form, it is entirely observable and can be audited later. The problems of subjectivity which arise with human examiners are not as evident with a text-based, virtual examiner. (Having said which, because each student will have a different interaction, there is no escaping a degree of subjectivity arising from differing virtual examination question-and-answer sequences.)

An important aspect of this essay was that a slightly altered version of the essay was published (in 2014) and has been available in open access form since then. A naïve plagiarism detector can detect the similarity between the published paper and the essay, even though in this case there is no plagiarism involved. There is every reason to believe that the published paper is part of the corpus used to train LLMs. Happily, the LLM-based virtual examiner did not flag the essay up as potential plagiarism, but based its response on the interaction with the student. This suggests that the prompt to the LLM was sufficient to generate the desired simulation of an examination, and that the judgement of originality offered does not depend on a naïve similarity metric to existing text.

\section{Real-world deployment}

Any student who has submitted LLM-generated work for assessment will have an obvious route to surviving an LLM-based virtual examination: using the LLM to generate responses to be copied-and-pasted into the virtual examination chat. To mitigate this risk, submitted work would need to be virtually examined in an invigilated setting, with a secure computing environment for each student, and with invigilators ensuring examination conditions pertain. Some universities already provide such examination rooms with secure computing resources for examinations. These would also allow the assessors to conveniently provide for the transcripts to be visible to the invigilators in real-time, and to collect and pass the transcripts back to the assessors.

The usual risks with LLMs would be apparent, and steps would need to be taken to ensure that attempts to cheat the virtual examination could be detected. For example, students might be tempted to include in their submitted work secret instructions to the LLM examiner, as has been observed in publications submitted for review. Minimally, the virtual examination environment would need to accept the submitted work in a secure format, perhaps even plain text, and be designed to identify and highlight attempts to subvert the examination.

Providing transcripts to assessors would be essential to allowing human judgement of the originality of the work and the command of the subject by the student. Assessors would need to be aware of the fact that each student's virtual examination would be unique, and that it would be difficult to ensure that each student's examination was of equal difficulty. Other equity issues with \textit{viva voce} examinations would need to be considered, such as the impact on neurodivergent students.

One of the main advantages of the proposed virtual examination is its scalability, and the simultaneous resolution of one substantive problem with human \textit{viva voce} examinations: that of differing conditions and human biases creating certain forms of subjectivity in assessment. Because an LLM-based virtual examination can be administered in parallel, and this parallelisation has not intrinsic limit, an entire cohort of students can be assessed simultaneously (as long as the earlier point around invigilation addresses differences of examination conditions). Instead of only examining certain students \textit{viva voce}, every student submitting assessed work can be examined virtually as part of the submission process.

One of the authors, at least, can recall that as recently as 15 years ago, assessed work had to be submitted in person and on paper at a particular University building. It does not seem, therefore, that for assessed work, a return to a form of in-person submission inherently poses an unbearable burden, although it would involve a certain loss of the flexibility students have come to expect in recent years with online submission from home, and so forth.

\section{Conclusion}

The proposed virtual or simulated \textit{viva voce} examination is an example of using LLM-based tools to support student learning and assessment. The widespread availability of generative AI tools and their use by students has rapidly changed the assessment landscape, and this has particularly affected subjects where long-form assessed work has typically formed part of summative assessment. Our proposal offers a way to use technology to not only mitigate the problems arising from generative AI tools, but to potentially improve upon existing interactive oral examinations by systematising and regularising the examination process. Developing and deploying a virtual examiner is relatively straightforward, and so institutions could implement this idea at minimal cost and may even be able to use LLM-based examinations to improve on non-interactive assessments.

\printbibliography

\end{document}